\documentclass[
 reprint,
 superscriptaddress,
 amsmath,amssymb,
 aps,
 prb, 
 longbibliography 
]{revtex4-2}

\usepackage{graphicx}
\usepackage{dcolumn}
\usepackage{bm}
\usepackage[dvipsnames]{xcolor}
\usepackage{hyperref}
\hypersetup{colorlinks=true, linkcolor=blue, citecolor=blue, urlcolor=blue}
\allowdisplaybreaks
\renewcommand{\vec}[1]{\mathbf{#1}}
\newcommand{\uvec}[1]{\hat{\mathbf{#1}}}
\renewcommand{\textcolor}[2]{#2}

\begin{document}

\title{Circular Huygens Dipoles: Unidirectional Spin-Angular Momentum from Achiral Nanoparticles}

\author{Esmaeel Zanganeh}
\email{esmaeelzanganeh@gmail.com}
\affiliation{London Centre for Nanotechnology, 19 Gordon St, London, WC1H 0AH, United Kingdom}

\author{Antonio Lombardo}
\affiliation{London Centre for Nanotechnology, 19 Gordon St, London, WC1H 0AH, United Kingdom}
\affiliation{Department of Electronic $\&$ Electrical Engineering, University College London, Malet Place, London, WC1E 7JE, United Kingdom}

\begin{abstract}
Simultaneous control over the directionality and spin of light at the nanoscale is a central goal in nanophotonics with applications ranging from quantum information to advanced biosensing. We introduce the concept of the Circular Huygens Dipole and numerically demonstrate its realization in a single Si nanocuboid. We show that the polarization of an incident linear wave controls the interference between co-located circular electric and magnetic dipoles excited in phase quadrature. This enables deterministic switching of the forward-scattered radiation between purely right- and left-circularly polarized states. The system also functions as a directional spin-to-linear polarization converter. Our findings establish a robust, passive method for reconfigurable spin-directional control in a simple, monolithic silicon nanostructure, opening avenues for chip-scale spin-optics, chiral quantum interfaces, and novel sensing platforms.
\end{abstract}

\maketitle

\section{Introduction}

Controlling the fundamental properties of light—its momentum and spin—is a cornerstone of modern photonics \cite{Forbes2021, Wang2012, Nicolas2014}. A primary goal is to master both simultaneously, enabling the routing of light along a chosen path with a specific helicity. This interplay is central to emerging fields like chiral quantum optics \cite{Lodahl2017, Gonzalez-Tudela2024} and topological photonics \cite{Barik2018}, and is crucial for engineering photonic spin-orbit interactions \cite{Bliokh2015, Guo2019, Solner2015}.

Achieving such control requires merging two key concepts. The first is momentum control via unidirectional scattering, epitomized by the Huygens dipole, where interfering electric and magnetic multipoles cancel radiation in one direction \cite{Kerker1983, Geffrin2012, Fu2013, Liu2018}. \textcolor{blue}{While this achieves directionality, its emission is strictly linearly polarized (LP).} The second \textcolor{blue}{key concept} is spin control, which originates from two fundamental source types. \textcolor{blue}{A \textbf{circular dipole} is composed of two orthogonal linear dipoles of the same nature oscillating in phase quadrature  (e.g., an electric dipole $p_x \pm i p_y$, or a magnetic dipole $m_x \pm i m_y$). Through spin-momentum locking \cite{Bliokh2015, Lodahl2017}, its evanescent near-fields can unidirectionally excite guided modes along a material interface \cite{Rodriguez2013}. In the far-field, however, the polarization of its radiation varies spatially from purely circular along its normal axis to purely linear in its transverse plane \cite{Cheng2023, Shi2023}. This is fundamentally distinct from a \textbf{chiral dipole}, which consists of parallel electric and magnetic moments oscillating in phase quadrature (e.g., $p_{x}\pm im_{x}$). The chiral dipole represents the fundamental eigenmode of a structurally chiral particle. While it radiates purely circularly polarized (CP) light across its entire toroidal emission profile \cite{Eismann2018, Wozniak2019}, it is incapable of unidirectional emission. Thus, neither fundamental spin source can achieve free-space directionality. Consequently, momentum and spin control remain largely decoupled.}

\textcolor{blue}{Existing strategies to bridge this gap often involve significant complexity. Structurally chiral nanoparticles can link spin and direction, but their response is fixed by their static geometry \cite{Xie2025}. Geometrically simple nanoparticles can be used, but they demand complex illumination schemes. For instance, directional scattering has been achieved by driving a spinning magnetic dipole alongside an orthogonal linear electric dipole; however, this necessitates tightly focused vector vortex beams combining radial polarization with azimuthal phase vortices, and the resulting scattered light in the transverse plane remains LP \cite{Wei2017}. Similarly, generating well-defined scattered helicity from simple lossy spheres requires precise interferometric pumping using multiple dephased, counter-propagating CP plane waves \cite{Carretero2022}. While CP scattering can be generated from an achiral sphere using a single LP plane wave, the emission is quadridirectional rather than unidirectional and depends critically on the surrounding medium \cite{Negoro2024}.}

\textcolor{blue}{A platform capable of overcoming these limitations to generate purely unidirectional, single-helicity CP radiation from a single achiral scatterer under simple plane-wave illumination has remained elusive. In this work, we address this void by demonstrating that the requisite phase and amplitude transformations can be embedded directly into the structural anisotropy of a monolithic silicon nanocuboid. Under normal incidence from a single LP plane wave, the particle's engineered dimensions simultaneously co-excite a circular electric dipole and a circular magnetic dipole in precise phase quadrature. This realizes a Circular Huygens Dipole, a fundamentally new source that emits a single-helicity, purely CP wave strictly forward along the optical axis. The emitted spin state can be deterministically switched between pure right- and left-circularly polarized states simply by rotating the polarization of the incident wave.}

\begin{figure*}[t!]
\includegraphics[width=0.95\textwidth]{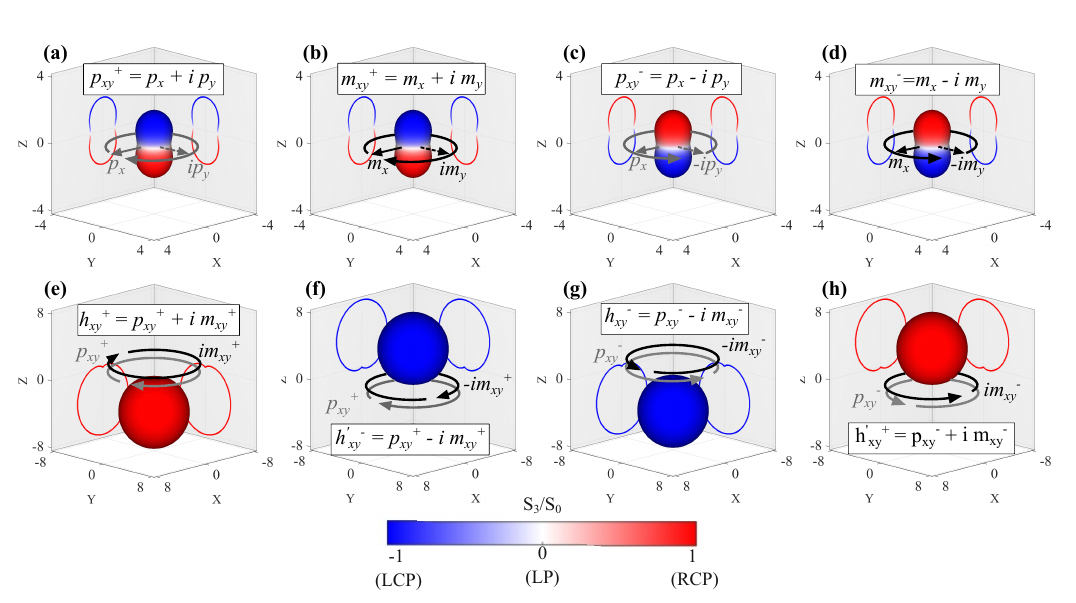} 
\caption{\textbf{Principle of the Circular Huygens Dipole.} 
3D far-field radiation patterns for ideal point dipoles. Radius is proportional to intensity ($S_0$) and color indicates the degree of circular polarization ($S_3/S_0$). Stokes parameters are $S_0 = |E_\theta|^2 + |E_\phi|^2$ and $S_3 = 2\,\mathrm{Im}(E_\theta E_\phi^*)$. Handedness is defined from the source viewpoint.
(a)-(d) Bidirectional radiation from fundamental right-handed ($p_{xy}^+, m_{xy}^+$) and left-handed ($p_{xy}^-, m_{xy}^-$) circular dipoles. 
(e)–(h) Their interference yields four unique unidirectional spin states: (e) Forward RCP from $h_{xy}^+=p_{xy}^+ + i m_{xy}^+$. (f) Backward LCP from $h'_{xy}{}^-=p_{xy}^+ - i m_{xy}^+$. (g) Forward LCP from $h_{xy}^-=p_{xy}^- - i m_{xy}^-$. (h) Backward RCP from $h'_{xy}{}^+=p_{xy}^- + i m_{xy}^-$.
}
\label{fig1}
\end{figure*}

\section{Principle of the Circular Huygens Dipole}
The principle of the Circular Huygens Dipole is built from the interference of fundamental circular electric dipoles, $p_{xy}^{\pm} = p_x \pm i p_y$, and magnetic dipoles, $m_{xy}^{\pm} = m_x \pm i m_y$, as illustrated in Fig.~\ref{fig1}. We define forward and backward propagation along the $-z$ and $+z$ axes, respectively. As shown in Figs.~\ref{fig1}(a-d), \textcolor{blue}{unlike a Huygens source, these elementary circular dipoles do not radiate unidirectionally.} Their far-zone electric fields are given by:
\begin{align}
 \vec{E}_{p_{xy}^{\pm}} &= C_E(r) e^{\pm i\phi}(\cos\theta\hat{\theta} \pm i\hat{\phi}) \label{eq:E_p_circ} \\
 \vec{E}_{m_{xy}^{\pm}} &= C_M(r) e^{\pm i\phi}(\pm i\hat{\theta} - \cos\theta\hat{\phi}) \label{eq:E_m_circ}
\end{align}
where $C_E(r) = k^2 p_0 (4\pi\epsilon_0 r)^{-1} e^{ikr}$ and $C_M(r) = k^2 Z_0 m_0 (4\pi r)^{-1} e^{ikr}$ contain the dipole moment magnitudes ($p_0$, $m_0$) and spatial dependence. These bidirectional fields are a manifestation of spin-momentum locking and can be decomposed into the sum of two oppositely-directed, oppositely-polarized cardioid patterns. For the right-handed (+) sources, this decomposition is (see Appendix~\ref{app:theory} for full derivation):
\begin{widetext}
\begin{align}
 \vec{E}_{p_{xy}^{+}} &= C_E(r)\left[ \underbrace{\frac{1}{2} e^{i\phi} (1-\cos\theta)(-\uvec{\theta} + i\uvec{\phi})}_{\text{Forward RCP Cardioid}} + \underbrace{\frac{1}{2} e^{i\phi} (1+\cos\theta)(\uvec{\theta} + i\uvec{\phi})}_{\text{Backward LCP Cardioid}} \right] \label{eq:p_decomp} \\
 \vec{E}_{m_{xy}^{+}} &= C_M(r)\left[ \underbrace{-\frac{i}{2} e^{i\phi} (1-\cos\theta)(-\uvec{\theta} + i\uvec{\phi})}_{\text{Forward RCP Cardioid}} + \underbrace{\frac{i}{2} e^{i\phi} (1+\cos\theta)(\uvec{\theta} + i\uvec{\phi})}_{\text{Backward LCP Cardioid}} \right] \label{eq:m_decomp}
\end{align}
\end{widetext}
An analogous decomposition exists for the left-handed ($-$) sources, where they radiate a forward LCP cardioid and a backward RCP cardioid.

The key to unidirectional emission lies in the intrinsic, handedness-dependent phase relationship between the far-fields radiated by the circular electric and magnetic dipoles. For the right-handed (+) source, comparing the components in Eqs.~(\ref{eq:p_decomp}) and (\ref{eq:m_decomp}) shows that the far-field from the magnetic dipole ($\vec{E}_m$) lags that from the electric dipole ($\vec{E}_p$) by 90° in the forward direction but leads by 90° in the backward direction. For the left-handed (-) source, this behavior is reversed: the field from the magnetic dipole leads forward and lags backward. This inherent, direction-dependent phase flip is the physical origin of the Huygens and anti-Huygens conditions. By introducing an appropriate external quadrature phase shift between the source moments, we can engineer perfect constructive or destructive interference in a chosen direction and handedness.

\subsection{Huygens Dipoles (Forward Emission).} By enforcing that the fields radiated by the electric and magnetic dipoles have equal amplitudes (the impedance matching condition, $|p_0|=|m_0|/c$), perfect interference can be achieved, where $C_E(r)=C_M(r)=C_0(r)$. A \textit{Right-Handed Circular Huygens Dipole} ($h_{xy}^{+} = p_{xy}^{+} + i m_{xy}^{+}$, Fig.~\ref{fig1}e) is formed when $m_{xy}^{+}$ leads $p_{xy}^{+}$ by $90^{\circ}$. This external lead cancels the intrinsic forward lag for constructive interference and adds to the intrinsic backward lead for destructive interference, yielding a purely forward-propagating RCP beam:
\begin{equation}
 \vec{E}_{h_{xy}^+} = C_0(r) e^{i\phi} (1 - \cos\theta) (-\hat{\theta} + i\hat{\phi}).
\end{equation}

A \textit{Left-Handed Circular Huygens Dipole} ($h_{xy}^{-} = p_{xy}^{-} - i m_{xy}^{-}$, Fig.~\ref{fig1}g), where $m_{xy}^{-}$ lags $p_{xy}^{-}$, produces forward-propagating LCP:
\begin{equation}
 \vec{E}_{h_{xy}^-} = - C_0(r) e^{-i\phi} (1 - \cos\theta) (\hat{\theta} + i\hat{\phi}).
\end{equation}

\subsection{Anti-Huygens Dipoles (Backward Emission).} When a right-handed source $p_{xy}^{+}$ interferes with a lagging $m_{xy}^{+}$ ($h'_{xy}{}^{-} = p_{xy}^{+} - i m_{xy}^{+}$, Fig.~\ref{fig1}f), the interference pattern is reversed, yielding a backward-propagating LCP beam:
\begin{equation}
 \vec{E}_{h'_{xy}{}^{-}} = C_0(r) e^{i\phi}(1+\cos\theta)(\hat{\theta} + i\hat{\phi}).
\end{equation}
Similarly, for a left-handed source with a leading magnetic moment ($h'_{xy}{}^{+} = p_{xy}^{-} + i m_{xy}^{-}$, Fig.~\ref{fig1}h), the emission is backward RCP:
\begin{equation}
 \vec{E}_{h'_{xy}{}^{+}} =- C_0(r) e^{-i\phi}(1+\cos\theta)(-\uvec{\theta} + i\uvec{\phi}).
\end{equation}
These four unidirectional single-helicity fields (Eqs. (5-8) and Fig.~\ref{fig1}e-h) represent eigenvectors of the helicity operator \cite{Vavilin2024, Nikitina2024}. Full derivations for all ideal dipole fields are provided in Appendix~\ref{app:theory}.

\begin{figure}[b!]
\includegraphics[width=\linewidth]{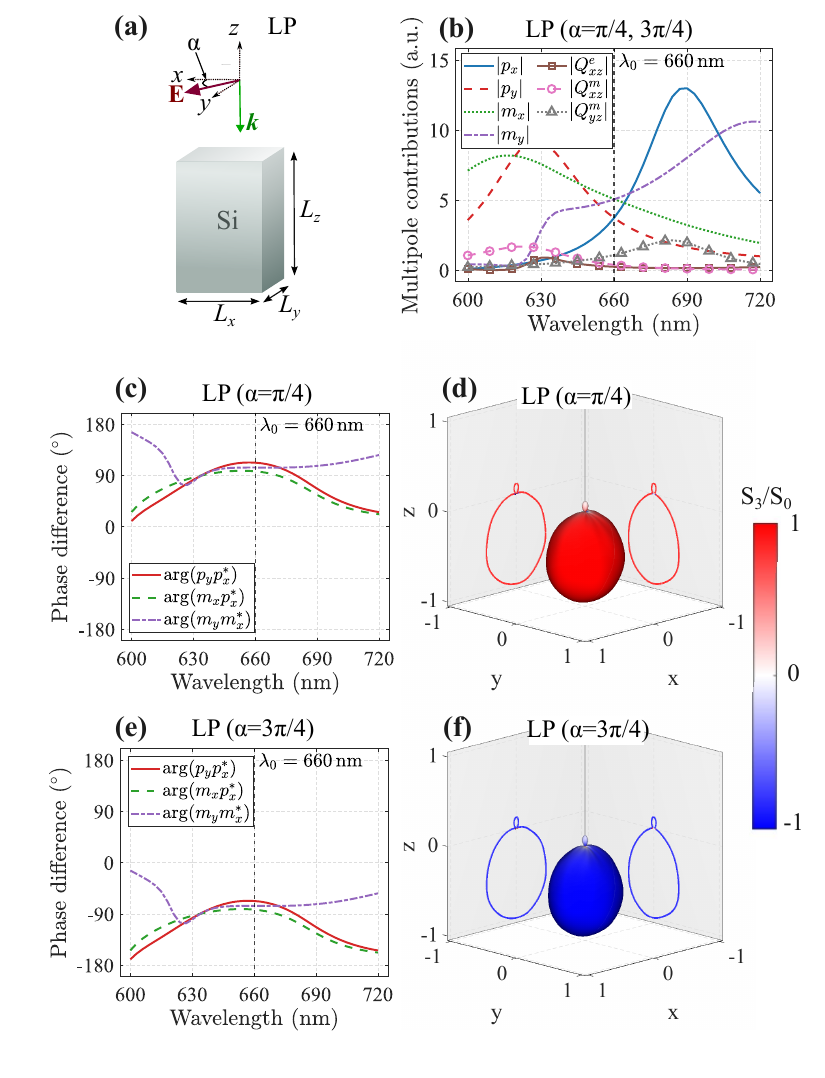}
\caption{\textbf{Realization of Circular Huygens Dipoles in a Si Nanocuboid.} 
(a) Scattering geometry: Si nanocuboid ($L_x=128$, $L_y=94$, $L_z=440$ nm) under LP illumination at angle $\alpha$. 
(b) Multipole magnitudes, identical for $\alpha=\pi/4$ and $3\pi/4$, show comparable strength of $p_x,p_y,m_x,m_y$ at $\lambda_0=660$ nm. 
(c, d) For $\alpha=\pi/4$, phase differences of $\approx+90^\circ$ excite $p_{xy}^+$ and $m_{xy}^+$, realizing a Right-Handed Circular Huygens Dipole ($h_{xy}^+$) and producing forward RCP scattering. 
(e, f) For $\alpha=3\pi/4$, phase differences flip to $\approx-90^\circ$, exciting $p_{xy}^-$ and $m_{xy}^-$ and yielding forward LCP scattering.}
\label{fig2}
\end{figure}

\begin{figure}[t!]
\includegraphics[width=0.95\linewidth]{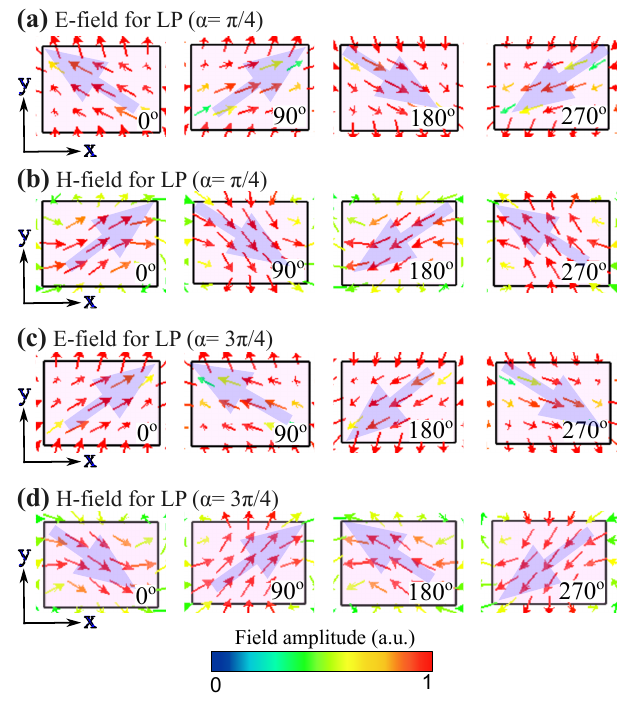}
\caption{\textbf{Near-Field Origin of Circular Huygens Dipoles.} 
E- and H-field distributions in the central xy-plane at $\lambda_0=660$ nm. \textcolor{blue}{The overlaid translucent arrows indicate the instantaneous direction of the net macroscopic dipole moment.} 
(a, b) For $\alpha=\pi/4$, fields co-rotate clockwise (CW), exciting $p_{xy}^+$ and $m_{xy}^+$. The magnetic field leads the electric by $\approx+90^\circ$, realizing $h_{xy}^+ = p_{xy}^+ + i m_{xy}^+$. 
(c, d) For $\alpha=3\pi/4$, fields co-rotate counter-clockwise (CCW), exciting $p_{xy}^-$ and $m_{xy}^-$. The magnetic field lags by $\approx-90^\circ$, realizing $h_{xy}^- = p_{xy}^- - i m_{xy}^-$. 
These near-fields directly show how incident polarization controls dipole handedness and interferometric phase.}
\label{fig3}
\end{figure}

\section{RESULTS AND DISCUSSION}
\textcolor{blue}{To physically realize the Circular Huygens Dipole,} we perform full-wave numerical simulations \textcolor{blue}{of the scattering from a silicon nanocuboid (Fig.~\ref{fig2}a)} using the COMSOL Multiphysics Frequency domain solver. The particle ($L_x=128$, $L_y=94$, $L_z=440$ nm) is modeled using the experimental refractive index of crystalline silicon \cite{Palik1998_SM} and is illuminated by a \textcolor{blue}{normally incident LP} plane wave propagating along the $-z$ direction. The spherical air domain is terminated by perfectly matched layers (PMLs). To visualize the 3D radiation patterns, the total far-field intensity ($S_0$) is plotted as the surface radius and colored by the normalized third Stokes parameter ($S_3/S_0$). The resulting \textcolor{blue}{internal} near-fields are then used to calculate the multipole moments \cite{Evlyukhin2019, zanganeh2021anapole, Zanganeh2021Non, Basharin}, as detailed in Appendix~\ref{app:formulas}.
 
\textcolor{blue}{Before analyzing the spectral response, it is vital to establish the structural properties and the physical intuition governing this configuration. Geometrically, the silicon nanocuboid possesses three mutually orthogonal spatial mirror symmetry planes ($x=0$, $y=0$, and $z=0$). These mirror symmetries strictly define the scatterer as an achiral object, precluding any intrinsic optical activity or circular dichroism. The generation of pure spin-angular momentum arises dynamically from the interplay between the particle's engineered structural anisotropy ($L_x \neq L_y$) and the projected linear excitation. When illuminated by a diagonal linear polarization ($\alpha = \pi/4$ or $3\pi/4$), the incident field projects equally onto the orthogonal fundamental axes. In a symmetric cross-section ($L_x = L_y$), the induced orthogonal dipoles would share identical resonance wavelengths. However, breaking this symmetry ($L_x \neq L_y$) imposes distinct localized boundary conditions along the $x$- and $y$-axes, spectrally splitting their fundamental resonance profiles. By simultaneously optimizing the lengths ($L_x, L_y, L_z$) relative to the target operational wavelength ($\lambda_0 = 660\text{ nm}$), the fundamental Cartesian resonance profiles are spectrally separated just enough to exploit their phase dispersions. At $\lambda_0$ under diagonal excitation ($\alpha = \pi/4$), this geometric detuning induces a precise $+90^\circ$ phase shift between the orthogonal electric dipole components ($p_x$ and $p_y$) and a simultaneous $+90^\circ$ shift between the magnetic dipole components ($m_x$ and $m_y$), dynamically synthesizing co-located right-handed circular dipoles. Crucially, because the structural anisotropy imposes a fixed phase difference between the orthogonal physical axes, rotating the incident linear polarization by $90^\circ$ (to $\alpha = 3\pi/4$) perfectly flips the sign of the incident field projection along one of these axes. This inverts the total relative phase delay from $+90^\circ$ to $-90^\circ$, flipping both the electric and magnetic dipoles to a left-handed state and seamlessly switching the forward emission from right- to left-circularly polarized. The spin state of the emitted radiation is therefore not a static intrinsic property of the structure, but a dynamically synthesized state controlled deterministically by the polarization of the incident linear plane wave.}
 
Figure 2b shows the multipole scattering magnitudes, which are identical for both diagonal polarizations ($\alpha=\pi/4, 3\pi/4$). The spectrum reveals an optimal wavelength of $\lambda_0=660\text{ nm}$, where the four required Cartesian dipole moments ($p_x, p_y, m_x, m_y$) are strongly co-excited with comparable magnitudes.

For an incident polarization $\alpha=\pi/4$, the nanoparticle's anisotropic response induces the key phase relationships shown in Fig.~\ref{fig2}c. Specifically, the electric dipole component $p_y$ leads $p_x$ by $\approx+90^\circ$, creating a right-handed circular electric dipole ($p_{xy}^{+} = p_x+ip_y$). Likewise, $m_y$ leads $m_x$ by $\approx+90^\circ$, forming a co-located right-handed circular magnetic dipole ($m_{xy}^{+} = m_x+im_y$). Critically, the magnetic dipole simultaneously leads the electric dipole by $\approx+90^\circ$ (e.g., $\arg(m_xp_x^*) \approx +90^\circ$). This fulfills the precise interferometric requirement for a Right-Handed Circular Huygens Dipole ($h_{xy}^{+} = p_{xy}^{+} + i m_{xy}^{+}$). The resulting constructive interference produces the forward (-z) RCP radiation shown in Fig.~\ref{fig2}d, which matches the ideal model in Fig.~\ref{fig1}(e).

Conversely, for the orthogonal polarization $\alpha=3\pi/4$, all relative phases flip to $\approx-90^\circ$, as depicted in Fig.~\ref{fig2}e. This sign reversal means the electric dipole component $p_y$ now lags $p_x$, creating a left-handed circular electric dipole ($p_{xy}^- = p_x - ip_y$), while $m_y$ lags $m_x$, forming a co-located left-handed circular magnetic dipole ($m_{xy}^- = m_x - im_y$). Critically, the magnetic dipole itself now lags the electric dipole, perfectly satisfying the interferometric requirement for a Left-Handed Circular Huygens Dipole ($h_{xy}^- = p_{xy}^- - i m_{xy}^-$). This produces the forward-propagating LCP radiation shown in Fig.~\ref{fig2}f, which corresponds to the ideal case in Fig.~\ref{fig1}g.
\begin{figure}[t!]
\includegraphics[width=\linewidth]{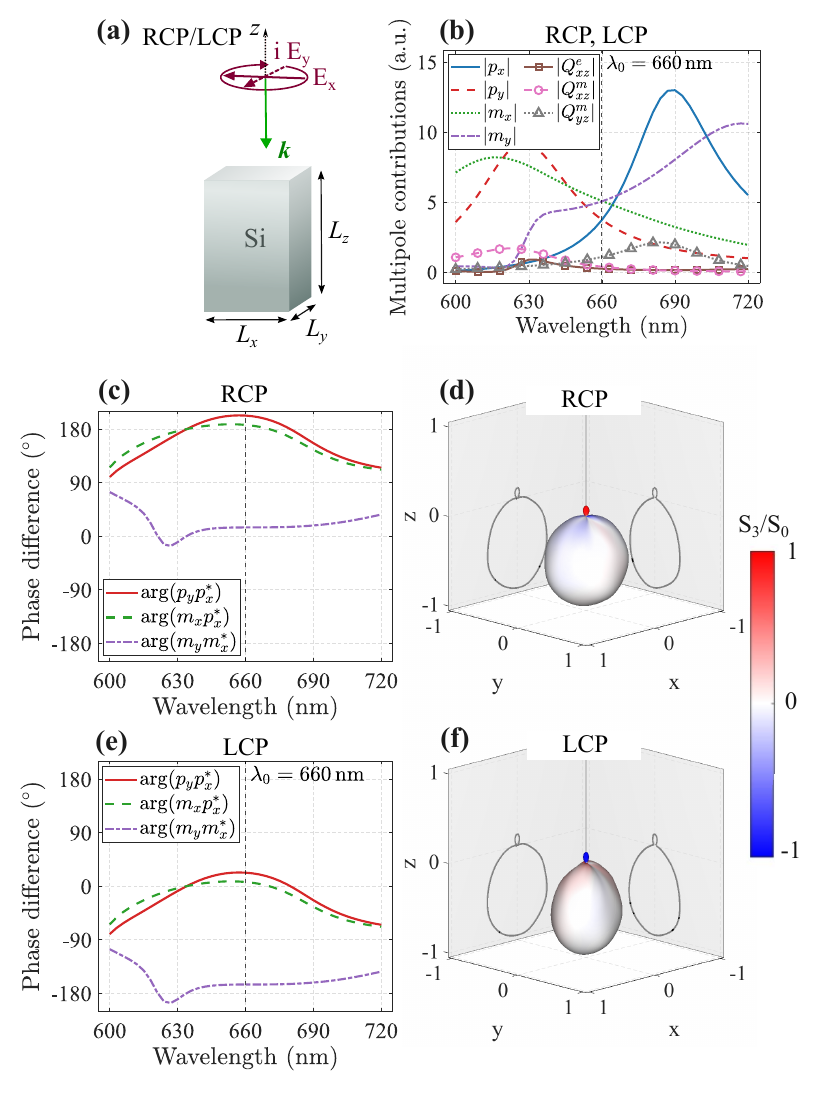}
\caption{\textbf{Directional Linear Scattering from Circular Excitation.} 
(a) A Si nanocuboid under normally incident CP illumination. 
(b) Excited multipole magnitudes are identical to the linear case (Fig.~2b). 
(c, d) For RCP input, phased orthogonal electric and magnetic dipoles are excited that satisfy the linear Huygens condition, producing forward (-z) LP radiation. 
(e, f) LCP input excites a complementary set of dipoles, again yielding directional linear scattering.}
\label{fig4}
\end{figure}

\begin{figure}[t!]
\includegraphics[width=\linewidth]{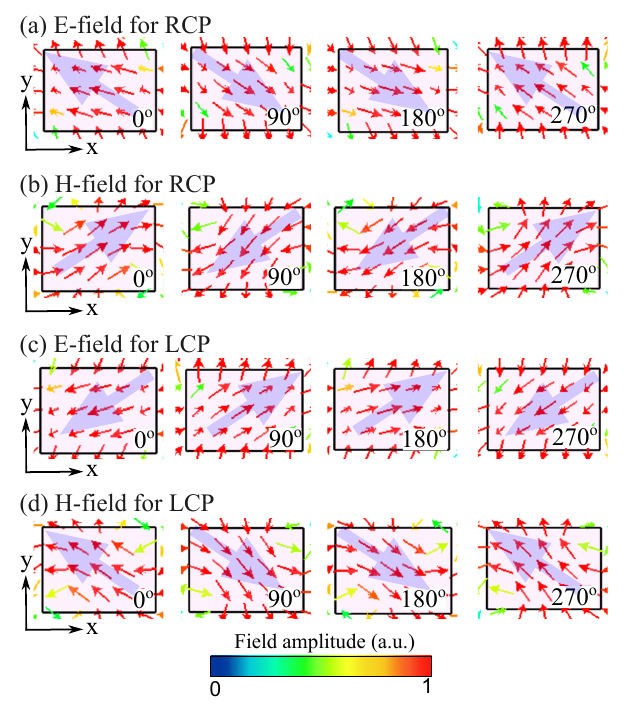}
\caption{\textbf{Near-Field Visualization of Circular-to-Linear Polarization Conversion.} Internal E- and H-field distributions in the central xy-plane of the nanocuboid at $\lambda_0=660$~nm for CP incidence. \textcolor{blue}{The overlaid translucent arrows indicate the instantaneous direction of the net macroscopic dipole moment, clearly highlighting the linear oscillation axes.} (a, b) For incident RCP light, the E-field and H-field exhibit linear oscillations along the orthogonal diagonals $y=-x$ and $y=x$, respectively. (c, d) For incident LCP light, the oscillation axes are swapped.}
\label{fig5}
\end{figure}
The physical origin of these circular dipoles is evident in the internal near-fields. Figure~\ref{fig3} shows that for $\alpha=\pi/4$, both the electric and magnetic fields are driven into a strong clockwise (CW) rotation (Fig.~\ref{fig3}a,b), generating the right-handed dipoles \textcolor{blue}{(See Supplemental Material \cite{SM_Videos} for a time-harmonic animation of this near-field rotation under linear excitation)}. The magnetic field rotation leads the electric field by $\approx+90^\circ$, directly confirming the realization of $h_{xy}^{+}$. For $\alpha=3\pi/4$, the fields co-rotate counter-clockwise (CCW) (Fig.~\ref{fig3}c,d), creating left-handed circular dipoles where the magnetic field lags, fulfilling the condition for $h_{xy}^{-}$.

The nanocuboid also exhibits a complementary wave-transforming capability (Fig.~\ref{fig4}). When the excitation is switched to a normally incident CP plane wave, the excited multipole magnitudes are identical to the LP case (Fig.~\ref{fig4}b). For incident RCP light, the phase analysis in Fig.~\ref{fig4}(c) reveals a remarkable transformation: the relation $\text{arg}(p_y p_x^*) \approx +180^\circ$ indicates an overall electric dipole oscillating linearly along the $y=-x$ diagonal, while $\text{arg}(m_y m_x^*) \approx 0^\circ$ indicates a magnetic dipole oscillating linearly along the orthogonal $y=x$ diagonal.

\begin{figure*}[t]
    \centering
    \includegraphics[width=\linewidth]{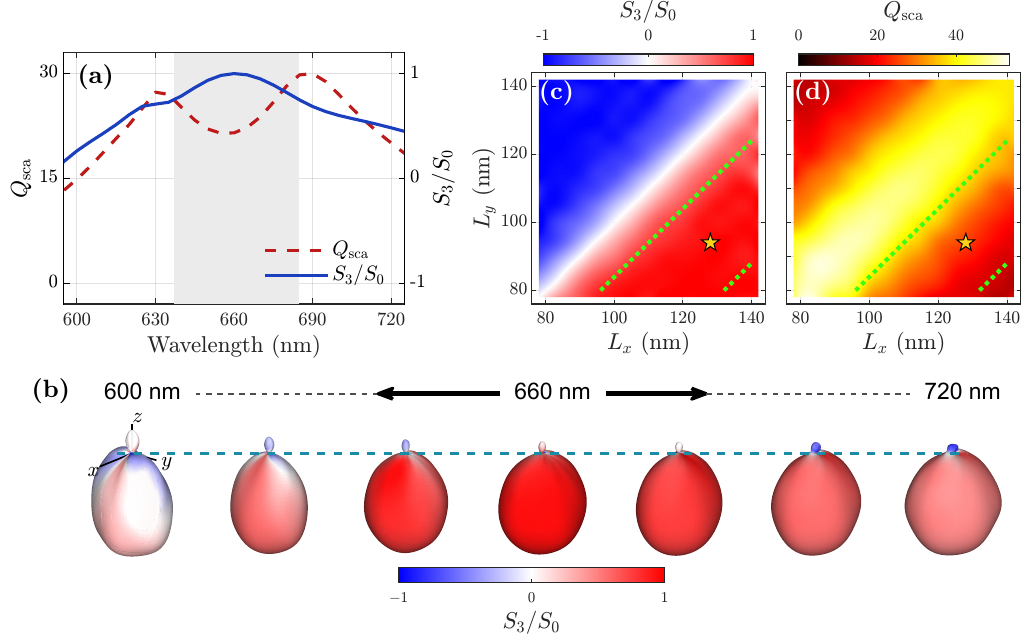}
    \caption{\textcolor{blue}{\textbf{Spectral bandwidth and structural robustness of the Circular Huygens Dipole.} \textbf{(a)} Wavelength-dependent scattering efficiency ($Q_{\text{sca}}$, dashed red curve) and forward circular polarization purity ($S_3/S_0$, solid blue curve) for the optimized Si nanocuboid under diagonal linear excitation ($\alpha = \pi/4$). The gray shaded region defines the high-performance operational window ($Q_{\text{sca}} > 20$ and $S_3/S_0 > 0.8$). \textbf{(b)} Spectral evolution of the 3D far-field radiation patterns from $\lambda = 600\text{ nm}$ to $720\text{ nm}$, color-coded by the local $S_3/S_0$ distribution. The solid black arrows indicate the broad spectral band where the forward-directed RCP cardioid is maintained. \textbf{(c, d)} 2D parametric mappings of (c) polarization purity and (d) scattering efficiency as functions of the orthogonal cross-sectional dimensions $L_x$ and $L_y$. The dotted green lines define the explicit $\pm 20\text{ nm}$ geometric tolerance bounds that satisfy the high-performance criteria, symmetrically flanking the optimal target geometry (gold star, $L_x = 128\text{ nm}$, $L_y = 94\text{ nm}$).}}
    \label{fig6}
\end{figure*}

This mechanism is confirmed by the internal near-fields, shown in Fig.~\ref{fig5} \textcolor{blue}{(and the corresponding time-harmonic animations provided in the Supplementary Material \cite{SM_Videos})}. For incident RCP light (Fig.~\ref{fig5}a,b), the internal electric field oscillates linearly along the $y = -x$ diagonal, while the magnetic field oscillates along the orthogonal $y = x$ diagonal. This confirms the creation of the two orthogonal linear dipoles identified in the multipole analysis. For incident LCP light (Fig.~\ref{fig5}c,d), the oscillation axes are swapped, creating a linear electric dipole along the $y = x$ axis and a linear magnetic dipole along the $y = -x$ axis (Fig.~\ref{fig4}e).

In both CP-incidence cases, the relative phasing between these two excited orthogonal linear dipoles satisfies the conventional linear Huygens condition, resulting in unidirectional forward scattering (Fig.~\ref{fig4}d, f) that is purely LP ($S_3/S_0 \approx 0$). The nanocuboid thus functions as a directional circular-to-linear polarization converter, transforming incident spin into a LP momentum state.

\textcolor{blue}{\subsection{Spectral Bandwidth, Symmetry Requirements, and Structural Robustness}}

\textcolor{blue}{To rigorously validate the operational stability of the Circular Huygens Dipole mechanism and address potential constraints stemming from realistic experimental conditions, we first evaluate its spectral bandwidth. Figures \ref{fig6}(a) and \ref{fig6}(b) illustrate the wavelength-dependent performance for our optimized target geometry ($L_x = 128\text{ nm}$, $L_y = 94\text{ nm}$, $L_z = 440\text{ nm}$) under fixed diagonal linear illumination ($\alpha = \pi/4$). As shown in Fig.~\ref{fig6}(a), while the circular polarization purity reaches near-unity at the exact target design wavelength ($|S_3/S_0| \rightarrow 1$ at $660\text{ nm}$), this precise quadrature phase matching is not an isolated spectral singularity. Instead, the device maintains a broad, high-performance operational bandwidth—highlighted by the gray shaded region—where high scattering efficiency ($Q_{\text{sca}} > 20$) and high circular polarization purity ($|S_3/S_0| > 0.8$) simultaneously overlap. The stability of this spectral window is further confirmed by the chronological far-field timeline in Fig.~\ref{fig6}(b). This timeline demonstrates that the radiation pattern remains strictly unidirectional across the spectrum. Furthermore, it maintains a nearly pure RCP state over a wide bandwidth ($640\text{ nm}$, $660\text{ nm}$, and $680\text{ nm}$), which covers almost the entire high-performance gray shaded region.}

\textcolor{blue}{Expanding beyond the single-geometry case to address fabrication robustness, we systematically evaluate the structural parameter space. We performed a full-wave 2D parametric sweep of the orthogonal cross-sectional dimensions $L_x$ and $L_y$, varying both from $80\text{ nm}$ to $140\text{ nm}$ in discrete steps. For every geometric configuration in this sweep, the nanocuboid was illuminated by a normally incident linearly polarized plane wave fixed at the same diagonal orientation ($\alpha = \pi/4$). Because altering the physical dimensions naturally shifts the spectral position of the fundamental resonances, we did not restrict this 2D map to a single static wavelength. Instead, for each $(L_x, L_y)$ configuration, we spectrally tracked the phase-matching condition and extracted the far-field scattering parameters at its respective optimal operational wavelength. Figures \ref{fig6}(c) and \ref{fig6}(d) illustrate the resulting computed 2D landscapes of the forward polarization purity ($S_3/S_0$) and total scattering cross-section ($Q_{\text{sca}}$), respectively.}

\textcolor{blue}{Along the pure degenerate diagonal where $L_x = L_y$ (visible as the central white zero-purity band in Fig.~\ref{fig6}(c)), the in-plane structural isotropy of the nanocuboid remains unbroken. This spatial degeneracy forces the orthogonal dipole components to share identical spectral phase responses, precluding the quadrature phase delay ($\pm 90^\circ$) required to synthesize spin. Consequently, the scattered field is restricted to a linear state ($S_3/S_0 = 0$). Moving away from this degenerate white band by introducing an intentional structural anisotropy ($L_x \neq L_y$) breaks the spatial degeneracy, effectively splitting the fundamental orthogonal eigenmodes and unlocking the high-purity circular polarization domains (deep red and blue regions where $|S_3/S_0| \rightarrow 1$). Crucially, the anti-symmetric distribution of these domains visually dictates the symmetry requirements of the device. Orienting the structural anisotropy such that $L_x > L_y$ yields pure right-handed emission (red region). Conversely, the inverse anisotropic orientation ($L_x < L_y$)—which is physically equivalent to a $90^\circ$ rotation of the achiral particle under fixed linear illumination—perfectly inverts the relative phase delay, yielding pure left-handed emission (blue region). This confirms that the scatterer is fundamentally achiral, granting deterministic access to both orthogonal spin states simply by selecting the axis of anisotropy.}

\textcolor{blue}{While maintaining perfect quadrature phase relies on the structural anisotropy ($\Delta L = 34\text{ nm}$), the unified performance domain comfortably accommodates absolute dimensional variations. As explicitly delineated by the dotted green boundary lines in Figs.~\ref{fig6}(c) and \ref{fig6}(d), the fabricated cross-sectional dimensions can deviate by up to $\pm 20\text{ nm}$ from the optimal target ($L_x = 128\text{ nm}$, $L_y = 94\text{ nm}$) while remaining securely within the high-performance regime ($S_3/S_0 > 0.8$ and $Q_{\text{sca}} > 20$). This proves that standard lithographic fabrication errors will not compromise the device, confirming that the Circular Huygens condition forms a robust, fabricable structural workspace.}

\textcolor{blue}{To benchmark the performance of the Circular Huygens Dipole against existing mechanisms, one must consider the trade-off between excitation complexity and the emitted spin-momentum state. For instance, scattering from high-index spheres illuminated by complex focused vector vortex beams \cite{Wei2017} can achieve highly tunable directional scattering, but the scattered field remains strictly linearly polarized ($S_3/S_0 \approx 0$). Alternatively, achieving pure directional helicity ($|S_3/S_0| \rightarrow 1$) from simple lossy spheres has been demonstrated, but this requires the precise interferometric alignment and tuning of multiple counter-propagating, dephased circularly polarized plane waves \cite{Carretero2022}. In contrast, our nanocuboid achieves an exceptional quantitative intersection of these properties—simultaneous high scattering efficiency ($Q_{\mathrm{sca}} > 20$), perfect circular polarization purity ($|S_3/S_0| \rightarrow 1$ at $\lambda_0 = 660\text{ nm}$), and strictly forward-directed emission with a Forward-to-Backward Ratio (FBR) exceeding 10 (10 dB)—using only a single linearly polarized plane wave. This represents a significant reduction in optical complexity for generating efficient, free-space directional spin.}


\section{Conclusion}

In conclusion, we have introduced the concept of the Circular Huygens Dipole and demonstrated its physical realization in a single Si nanocuboid. We have shown that the helicity of the forward-scattered radiation can be deterministically switched by rotating the incident wave polarization. Furthermore, we demonstrated the system's dual functionality as a directional spin-to-linear polarization converter. This work provides a clear blueprint for achieving robust spin-directional control with simple, high-index dielectric nanostructures, paving the way for ultra-compact, reconfigurable components for spin optics, integrated quantum circuits, and advanced metasurfaces.

\begin{acknowledgments}
E. Z. thanks Andrey Miroshnichenko for discussions.

\end{acknowledgments}

\appendix 

\section{Theoretical Framework of Ideal Dipoles}
\label{app:theory}

This appendix provides a pedagogical walkthrough of the analytical derivations for the far-field radiation patterns of the ideal dipole sources, starting from fundamental principles.

\begin{figure*}
\includegraphics[width=0.85\textwidth]{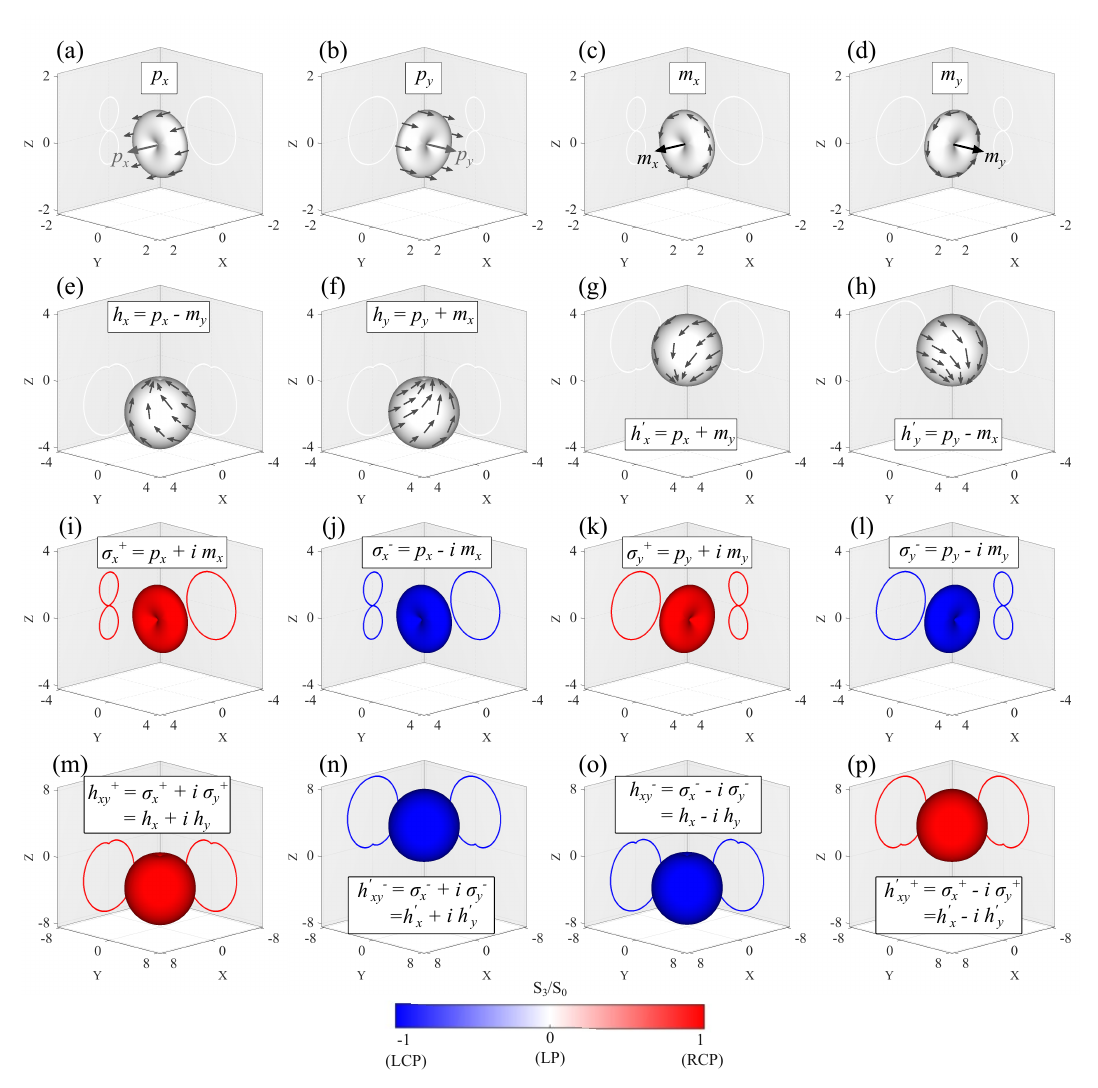}
\caption{\textbf{Alternative Pathways to Directional Spin Emission and Basis Equivalence.} (a)-(d) Radiation patterns of fundamental linear electric ($p_x, p_y$) and magnetic ($m_x, m_y$) dipoles. (e)-(h) Unidirectional, \textcolor{blue}{LP} radiation from linear Huygens (forward, e-f) and Anti-Huygens (backward, g-h) dipoles. (i)-(l) Radiation from chiral dipoles ($\sigma_x^\pm, \sigma_y^\pm$), which produce \textcolor{blue}{CP} toroidal patterns. (m)-(p) Synthesis of the four unidirectional, \textcolor{blue}{CP} sources from Fig.~1 of the main text, demonstrating that each can be constructed from a superposition of linear Huygens or chiral dipoles.}
\label{figS1}
\end{figure*}

\subsection{Fundamental Equations and Cartesian Basis}
We begin with the standard expressions for the far-zone electric field ($\vec{E}$) from time-harmonic ($e^{-i\omega t}$) point dipoles at the origin \cite{Jackson_SM}:
\begin{align}
\vec{E}_{p} &= \frac{k^2}{4\pi\epsilon_0} \frac{e^{ikr}}{r} (\uvec{n} \times \vec{p}) \times \uvec{n} \\
\vec{E}_{m} &= -\frac{k^2 Z_0}{4\pi} \frac{e^{ikr}}{r} (\uvec{n} \times \vec{m})
\end{align}
where $k=n\omega/c$ is the wavenumber in a medium of refractive index $n$, $c$ is the speed of light in vacuum, $\epsilon_0$ is the permittivity of free space, $Z_0$ is the impedance of free space, $r$ is the distance to the observation point, and $\uvec{n} = \vec{r}/r$ is the unit vector in the direction of observation. For clarity, we define the electric and magnetic field prefactors for dipoles with moment magnitudes $p_0$ and $m_0$:
\begin{align}
 C_E(r) = \frac{k^2 p_0}{4\pi\epsilon_0} \frac{e^{ikr}}{r} && \text{and} && C_M(r) = \frac{k^2 Z_0 m_0}{4\pi} \frac{e^{ikr}}{r}
\end{align}
The fields for linear dipoles along the x and y axes are then written as:
\begin{align}
 \vec{E}_{px} &= C_E(r) (\cos\theta\cos\phi\,\uvec{\theta} - \sin\phi\,\uvec{\phi}) \\
 \vec{E}_{py} &= C_E(r) (\cos\theta\sin\phi\,\uvec{\theta} + \cos\phi\,\uvec{\phi}) \\
 \vec{E}_{mx} &= -C_M(r) (\sin\phi\,\uvec{\theta} + \cos\theta\cos\phi\,\uvec{\phi})\\
 \vec{E}_{my} &= C_M(r) (\cos\phi\,\uvec{\theta} - \cos\theta\sin\phi\,\uvec{\phi})
\end{align}
These fundamental dipoles radiate with a toroidal intensity pattern and are \textcolor{blue}{LP}, as shown in Fig.~\ref{figS1}(a-d).
\subsection{Standard Far-Fields of Circular Dipoles}
The circular dipoles discussed in the main text are defined as superpositions of the linear dipoles from above. The detailed derivations of their far-fields are as follows:
\subsubsection{Right-Handed Electric Dipole ($p_{xy}^{+}$):} Defined as $p_{xy}^{+} = p_x + ip_y$, its radiated field is:
\begin{align}
\vec{E}_{p_{xy}^{+}} &= \vec{E}_{p_x} + i\vec{E}_{p_y} \nonumber\\
&= C_E(r) \big[ (\cos\theta\cos\phi\,\uvec{\theta} - \sin\phi\,\uvec{\phi}) \nonumber\\ 
& \qquad + i(\cos\theta\sin\phi\,\uvec{\theta} + \cos\phi\,\uvec{\phi}) \big] \nonumber\\
&= C_E(r) \big[ (\cos\theta\cos\phi + i\cos\theta\sin\phi)\,\uvec{\theta} \nonumber\\
& \qquad + (-\sin\phi+i\cos\phi)\,\uvec{\phi} \big] \nonumber\\
&= C_E(r) \left[ \cos\theta e^{i\phi}\,\uvec{\theta} + i e^{i\phi}\,\uvec{\phi} \right] \nonumber \\
&= C_E(r) e^{i\phi} (\cos\theta\,\uvec{\theta} + i\,\uvec{\phi})
\end{align}
\subsubsection{Left-Handed Electric Dipole ($p_{xy}^{-}$):} Defined as $p_{xy}^{-} = p_x - ip_y$, its radiated field is:
\begin{align}
\vec{E}_{p_{xy}^{-}} &= \vec{E}_{p_x} - i\vec{E}_{p_y} \nonumber\\
&= C_E(r) \big[ (\cos\theta\cos\phi\,\uvec{\theta} - \sin\phi\,\uvec{\phi}) \nonumber\\
& \qquad - i(\cos\theta\sin\phi\,\uvec{\theta} + \cos\phi\,\uvec{\phi}) \big] \nonumber\\
&= C_E(r) \big[ (\cos\theta\cos\phi - i\cos\theta\sin\phi)\,\uvec{\theta} \nonumber\\
& \qquad + (-\sin\phi-i\cos\phi)\,\uvec{\phi} \big] \nonumber\\
&= C_E(r) \left[ \cos\theta e^{-i\phi}\,\uvec{\theta} - i e^{-i\phi}\,\uvec{\phi} \right] \nonumber \\
&= C_E(r) e^{-i\phi} (\cos\theta\,\uvec{\theta} - i\,\uvec{\phi})
\end{align}
\subsubsection{Right-Handed Magnetic Dipole ($m_{xy}^{+}$):} Defined as $m_{xy}^{+} = m_x + im_y$, its radiated field is:
\begin{align}
\vec{E}_{m_{xy}^{+}} &= \vec{E}_{m_x} + i\vec{E}_{m_y} \nonumber\\
&= C_M(r) \big[ -(\sin\phi\,\uvec{\theta} + \cos\theta\cos\phi\,\uvec{\phi}) \nonumber\\
& \qquad + i(\cos\phi\,\uvec{\theta} - \cos\theta\sin\phi\,\uvec{\phi}) \big] \nonumber\\
&= C_M(r) \big[ (-\sin\phi + i\cos\phi)\,\uvec{\theta} \nonumber\\
& \qquad + (-\cos\theta\cos\phi - i\cos\theta\sin\phi)\,\uvec{\phi} \big] \nonumber\\
&= C_M(r) \left[ i e^{i\phi}\,\uvec{\theta} - \cos\theta e^{i\phi}\,\uvec{\phi} \right] \nonumber \\
&= C_M(r) e^{i\phi} (i\,\uvec{\theta} - \cos\theta\,\uvec{\phi})
\end{align}
\subsubsection{Left-Handed Magnetic Dipole ($m_{xy}^{-}$):} Defined as $m_{xy}^{-} = m_x - im_y$, its radiated field is:
\begin{align}
\vec{E}_{m_{xy}^{-}} &= \vec{E}_{m_x} - i\vec{E}_{m_y} \nonumber\\
&= C_M(r) \big[ -(\sin\phi\,\uvec{\theta} + \cos\theta\cos\phi\,\uvec{\phi}) \nonumber\\
& \qquad - i(\cos\phi\,\uvec{\theta} - \cos\theta\sin\phi\,\uvec{\phi}) \big] \nonumber\\
&= C_M(r) \big[ (-\sin\phi - i\cos\phi)\,\uvec{\theta} \nonumber\\
& \qquad + (-\cos\theta\cos\phi + i\cos\theta\sin\phi)\,\uvec{\phi} \big] \nonumber\\
&= C_M(r) \left[ -i e^{-i\phi}\,\uvec{\theta} - \cos\theta e^{-i\phi}\,\uvec{\phi} \right] \nonumber \\
&= C_M(r) e^{-i\phi} (-i\,\uvec{\theta} - \cos\theta\,\uvec{\phi})
\end{align}
These derivations yield the standard, bidirectional far-field expressions for right- (+) and left-handed (-) circular dipoles, shown in Fig.~1(a-d) of the main text:
\begin{align}
 \vec{E}_{p_{xy}^{\pm}} &= C_E(r) e^{\pm i\phi} (\cos\theta\,\uvec{\theta} \pm i\,\uvec{\phi}) \\
 \vec{E}_{m_{xy}^{\pm}} &= C_M(r) e^{\pm i\phi} (\pm i\,\uvec{\theta} - \cos\theta\,\uvec{\phi})
\end{align}
\subsection{Decomposition of Circular Dipole Far-Fields into Unidirectional Cardioids}
As stated in the main text, the bidirectional fields of elementary circular dipoles are a manifestation of spin-momentum locking. They can be decomposed into a sum of two oppositely-directed, oppositely-polarized cardioid patterns. The angular part of any \textcolor{blue}{CP} field can be written as a superposition of a forward-propagating cardioid with angular dependence $(1-\cos\theta)$ and a backward-propagating cardioid with angular dependence $(1+\cos\theta)$.
\subsubsection{Right-Handed (+) Sources:}
A right-handed source radiates RCP forward and LCP backward. The decomposition is as follows:
\begin{widetext}
\begin{align}
 \vec{E}_{p_{xy}^{+}} &= C_E(r) e^{i\phi}(\cos\theta\,\uvec{\theta} + i\,\uvec{\phi}) = \underbrace{ C_E(r) \left( \frac{1}{2} \right) e^{i\phi} (1-\cos\theta)(-\uvec{\theta} + i\uvec{\phi})}_{\text{Forward RCP Cardioid}} + \underbrace{ C_E(r) \left( \frac{1}{2} \right) e^{i\phi} (1+\cos\theta)(\uvec{\theta} + i\uvec{\phi})}_{\text{Backward LCP Cardioid}} \\
 \vec{E}_{m_{xy}^{+}} &= C_M(r) e^{i\phi}(i\,\uvec{\theta} - \cos\theta\,\uvec{\phi}) = \underbrace{ C_M(r) \left( -\frac{i}{2} \right) e^{i\phi} (1-\cos\theta)(-\uvec{\theta} + i\uvec{\phi})}_{\text{Forward RCP Cardioid}} + \underbrace{ C_M(r) \left( \frac{i}{2} \right) e^{i\phi} (1+\cos\theta)(\uvec{\theta} + i\uvec{\phi})}_{\text{Backward LCP Cardioid}}
\end{align}
\end{widetext}

\subsubsection{Left-Handed (-) Sources:}
A left-handed source radiates LCP forward and RCP backward. Following the same procedure, we find their decompositions:
\begin{widetext}
\begin{align}
 \vec{E}_{p_{xy}^{-}} &= C_E(r) e^{-i\phi}(\cos\theta\,\uvec{\theta} - i\,\uvec{\phi}) = \underbrace{ C_E(r) \left( -\frac{1}{2} \right) e^{-i\phi} (1-\cos\theta)(\uvec{\theta} + i\uvec{\phi})}_{\text{Forward LCP Cardioid}} + \underbrace{ C_E(r) \left( -\frac{1}{2} \right) e^{-i\phi} (1+\cos\theta)(-\uvec{\theta} + i\uvec{\phi})}_{\text{Backward RCP Cardioid}} \\
 \vec{E}_{m_{xy}^{-}} &= C_M(r) e^{-i\phi}(-i\,\uvec{\theta} - \cos\theta\,\uvec{\phi}) = \underbrace{ C_M(r) \left( -\frac{i}{2} \right) e^{-i\phi} (1-\cos\theta)(\uvec{\theta} + i\uvec{\phi})}_{\text{Forward LCP Cardioid}} + \underbrace{ C_M(r) \left( \frac{i}{2} \right) e^{-i\phi} (1+\cos\theta)(-\uvec{\theta} + i\uvec{\phi})}_{\text{Backward RCP Cardioid}}
\end{align}
\end{widetext}
These derivations provide the mathematical proof for the handedness-dependent phase relationships. By taking the ratio of the magnetic and electric prefactors, we find:
\begin{itemize}
 \item For right-handed (+) sources, the electric far-field from the magnetic dipole ($\vec{E}_m$) \textbf{lags} that from the electric dipole ($\vec{E}_p$) by 90° in the forward direction (phase ratio of $-i$) but \textbf{leads} by 90° in the backward direction (phase ratio of $+i$).
 \item For left-handed (-) sources, this behavior is reversed: the electric far-field from the magnetic dipole ($\vec{E}_m$) \textbf{leads} that from the electric dipole ($\vec{E}_p$) by 90° in the forward direction (phase ratio of $+i$) but \textbf{lags} by 90° in the backward direction (phase ratio of $-i$).
\end{itemize}

\subsection{Advanced Dipole Constructions}
To construct more complex radiation patterns, we enforce the impedance matching condition $|p_0| = |m_0|/c$. This makes the field amplitudes equal, i.e., $C_E(r) = C_M(r)$, since $Z_0/c = 1/\epsilon_0$. We denote this common prefactor as $C_0(r) = \frac{k^2 p_0}{4\pi\epsilon_0} \frac{e^{ikr}}{r}$.

\subsubsection{Linear Huygens and Anti-Huygens Dipoles:} Unidirectional, LP radiation is achieved by interfering orthogonal linear dipoles.
\begin{align}
 \vec{E}_{h_x} &= \vec{E}_{px} - \vec{E}_{my} = C_0(r) (\cos\theta - 1) (\cos\phi\,\uvec{\theta} - \sin\phi\,\uvec{\phi}) \\
 \vec{E}_{h_y} &= \vec{E}_{py} + \vec{E}_{mx} = C_0(r) (\cos\theta - 1) (\sin\phi\,\uvec{\theta} + \cos\phi\,\uvec{\phi}) \\
 \vec{E}_{h'_x} &= \vec{E}_{px} + \vec{E}_{my} = C_0(r) (1 + \cos\theta) (\cos\phi\,\uvec{\theta} - \sin\phi\,\uvec{\phi}) \\
 \vec{E}_{h'_y} &= \vec{E}_{py} - \vec{E}_{mx} = C_0(r) (1 + \cos\theta) (\sin\phi\,\uvec{\theta} + \cos\phi\,\uvec{\phi})
\end{align}
These sources produce unidirectional forward ($h_x, h_y$) and backward ($h'_x, h'_y$) radiation with linear polarization, as shown in Fig.~\ref{figS1}(e-h).

\subsubsection{Chiral Dipoles:} These sources, formed by parallel electric and magnetic dipoles ($\sigma_x^\pm = p_x \pm i m_x$), emit CP light in a toroidal pattern.
\begin{align}
\vec{E}_{\sigma_x^\pm} &= \vec{E}_{px} \pm i\vec{E}_{mx} = C_0(r) \left[ (\cos\theta\cos\phi \mp i\sin\phi)\uvec{\theta} \right. \nonumber \\
& \qquad \left. - (\sin\phi \pm i\cos\theta\cos\phi)\uvec{\phi} \right] \\
\vec{E}_{\sigma_y^\pm} &= \vec{E}_{py} \pm i\vec{E}_{my} = C_0(r) \left[ (\cos\theta\sin\phi \pm i\cos\phi)\uvec{\theta} \right. \nonumber \\
& \qquad \left. + (\cos\phi \mp i\cos\theta\sin\phi)\uvec{\phi} \right]
\end{align}
These sources produce toroidal radiation patterns with pure circular polarization, as shown in Fig.~\ref{figS1}(i-l).

\subsection{Circular Huygens and Anti-Huygens Dipoles:} 

To create unidirectional CP sources, we interfere the circular dipoles, again enforcing the impedance matching condition ($|p_0| = |m_0|/c$), which sets $C_E(r) = C_M(r) = C_0(r)$. This method starts with the decomposed forms of the circular electric and magnetic dipole fields and shows how the external phase shift ($\pm i$) leads to constructive interference in one direction and destructive interference in the other.

\subsubsection{Right-Handed Circular Huygens Dipole ($h_{xy}^{+} = p_{xy}^{+} + i m_{xy}^{+}$)}
The source field is defined by the superposition $\vec{E}_{h_{xy}^+} = \vec{E}_{p_{xy}^+} + i \vec{E}_{m_{xy}^+}$. Using the decomposed fields:
\begin{widetext}
\begin{align}
\vec{E}_{h_{xy}^+} = \vec{E}_{p_{xy}^+} + i \vec{E}_{m_{xy}^+} &= C_0(r) \left[ \frac{1}{2} e^{i\phi} (1-\cos\theta)(-\uvec{\theta} + i\uvec{\phi}) + \frac{1}{2} e^{i\phi} (1+\cos\theta)(\uvec{\theta} + i\uvec{\phi}) \right] \nonumber\\
& \qquad + i \cdot C_0(r) \left[ -\frac{i}{2} e^{i\phi} (1-\cos\theta)(-\uvec{\theta} + i\uvec{\phi}) + \frac{i}{2} e^{i\phi} (1+\cos\theta)(\uvec{\theta} + i\uvec{\phi}) \right] \nonumber\\
&= C_0(r) \left[  \frac{1}{2} e^{i\phi} (1-\cos\theta) +  \frac{1}{2} e^{i\phi} (1-\cos\theta) \right](-\uvec{\theta} + i\uvec{\phi}) \nonumber\\
& \qquad + C_0(r) \left[  \frac{1}{2} e^{i\phi} (1+\cos\theta) -  \frac{1}{2} e^{i\phi} (1+\cos\theta) \right](\uvec{\theta} + i\uvec{\phi}) \nonumber\\
&= C_0(r) e^{i\phi} (1 - \cos\theta) (-\uvec{\theta} + i\uvec{\phi})
\end{align}
\end{widetext}

\subsubsection{Left-Handed Circular Anti-Huygens Dipole ($h'_{xy}{}^{-} = p_{xy}^{+} - i m_{xy}^{+}$)}
The source field is $\vec{E}_{h'_{xy}{}^{-}} = \vec{E}_{p_{xy}^+} - i \vec{E}_{m_{xy}^+}$.
\begin{widetext}
\begin{align}
\vec{E}_{h'_{xy}{}^{-}} = \vec{E}_{p_{xy}^+} - i \vec{E}_{m_{xy}^+} &= C_0(r) \left[ \frac{1}{2} e^{i\phi} (1-\cos\theta)(-\uvec{\theta} + i\uvec{\phi}) + \frac{1}{2} e^{i\phi} (1+\cos\theta)(\uvec{\theta} + i\uvec{\phi}) \right] \nonumber\\
& \qquad - i \cdot C_0(r) \left[ -\frac{i}{2} e^{i\phi} (1-\cos\theta)(-\uvec{\theta} + i\uvec{\phi}) + \frac{i}{2} e^{i\phi} (1+\cos\theta)(\uvec{\theta} + i\uvec{\phi}) \right] \nonumber\\
&= C_0(r) \left[  \frac{1}{2} e^{i\phi} (1-\cos\theta) -  \frac{1}{2} e^{i\phi} (1-\cos\theta) \right](-\uvec{\theta} + i\uvec{\phi}) \nonumber\\
& \qquad + C_0(r) \left[  \frac{1}{2} e^{i\phi} (1+\cos\theta) +  \frac{1}{2} e^{i\phi} (1+\cos\theta) \right](\uvec{\theta} + i\uvec{\phi}) \nonumber\\
&= C_0(r) e^{i\phi} (1 + \cos\theta) (\uvec{\theta} + i\uvec{\phi})
\end{align}
\end{widetext}

\subsubsection{Left-Handed Circular Huygens Dipole ($h_{xy}^{-} = p_{xy}^{-} - i m_{xy}^{-}$)}
The source field is $\vec{E}_{h_{xy}^-} = \vec{E}_{p_{xy}^-} - i \vec{E}_{m_{xy}^-}$. We now use the decomposed forms for the left-handed dipoles.
\begin{widetext}
\begin{align}
\vec{E}_{h_{xy}^-} = \vec{E}_{p_{xy}^-} - i \vec{E}_{m_{xy}^-}  &= C_0(r) \left[  -\frac{1}{2} e^{-i\phi} (1-\cos\theta)(\uvec{\theta} + i\uvec{\phi}) -  \frac{1}{2} e^{-i\phi} (1+\cos\theta)(-\uvec{\theta} + i\uvec{\phi}) \right] \nonumber\\
& \qquad - i \cdot C_0(r) \left[ -\frac{i}{2} e^{-i\phi} (1-\cos\theta)(\uvec{\theta} + i\uvec{\phi}) + \frac{i}{2} e^{-i\phi} (1+\cos\theta)(-\uvec{\theta} + i\uvec{\phi}) \right] \nonumber\\
&= C_0(r) \left[  -\frac{1}{2} e^{-i\phi} (1-\cos\theta) -  \frac{1}{2} e^{-i\phi} (1-\cos\theta) \right](\uvec{\theta} + i\uvec{\phi}) \nonumber\\
& \qquad + C_0(r) \left[ - \frac{1}{2} e^{-i\phi} (1+\cos\theta) + \frac{1}{2} e^{-i\phi} (1+\cos\theta) \right](-\uvec{\theta} + i\uvec{\phi}) \nonumber\\
&= -C_0(r) e^{-i\phi} (1 - \cos\theta) (\uvec{\theta} + i\uvec{\phi})
\end{align}
\end{widetext}

\subsubsection{Right-Handed Circular Anti-Huygens Dipole ($h'_{xy}{}^{+} = p_{xy}^{-} + i m_{xy}^{-}$)}
The source field is $\vec{E}_{h'_{xy}{}^{+}} = \vec{E}_{p_{xy}^-} + i \vec{E}_{m_{xy}^-}$.
\begin{widetext}
\begin{align}
\vec{E}_{h'_{xy}{}^{+}}  = \vec{E}_{p_{xy}^-} + i \vec{E}_{m_{xy}^-}  &= C_0(r) \left[  -\frac{1}{2} e^{-i\phi} (1-\cos\theta)(\uvec{\theta} + i\uvec{\phi}) -  \frac{1}{2} e^{-i\phi} (1+\cos\theta)(-\uvec{\theta} + i\uvec{\phi}) \right] \nonumber\\
& \qquad + i \cdot C_0(r) \left[ -\frac{i}{2} e^{-i\phi} (1-\cos\theta)(\uvec{\theta} + i\uvec{\phi}) + \frac{i}{2} e^{-i\phi} (1+\cos\theta)(-\uvec{\theta} + i\uvec{\phi}) \right] \nonumber\\
&= C_0(r) \left[  -\frac{1}{2} e^{-i\phi} (1-\cos\theta) +  \frac{1}{2} e^{-i\phi} (1-\cos\theta) \right](\uvec{\theta} + i\uvec{\phi}) \nonumber\\
& \qquad + C_0(r) \left[  -\frac{1}{2} e^{-i\phi} (1+\cos\theta) - \frac{1}{2} e^{-i\phi} (1+\cos\theta) \right](-\uvec{\theta} + i\uvec{\phi}) \nonumber\\
&=- C_0(r) e^{-i\phi} (1 + \cos\theta) (-\uvec{\theta} + i\uvec{\phi})
\end{align}
\end{widetext}
These sources produce forward RCP ($h_{xy}^+$), backward LCP ($h'_{xy}{}^{-}$), forward LCP ($h_{xy}^{-}$), and backward RCP ($h'_{xy}{}^{+}$) radiation, respectively, as shown in Fig.~1(e-h) of the main text.

\subsection{Unifying Framework: The Equivalence of Dipole Bases}
A key insight, illustrated in Fig.~\ref{figS1}(m-p), is that any complex radiation pattern, including the four unidirectional spin states central to our work, is ultimately generated by a single, unique combination of the four fundamental linear dipoles ($p_x, p_y, m_x, m_y$). The different "pictures"—interfering circular dipoles, interfering linear Huygens dipoles, or interfering chiral dipoles—are not physically distinct phenomena. They are merely different, but equally valid, mathematical groupings of the same four fundamental sources. This provides a powerful, unifying framework for understanding complex emitters.

We demonstrate this explicitly for the Forward RCP state ($h_{xy}^{+}$). The required combination of linear moments is $p_x + i p_y + i m_x - m_y$. This single physical source can be interpreted in three ways:
\begin{itemize}
 \item \textbf{Circular Basis Viewpoint:} $(p_x + i p_y) + i (m_x + i m_y) \rightarrow p_{xy}^{+} + i m_{xy}^{+}$. 
 
 This is the intuitive picture used in the main text, where a right-handed electric dipole interferes with a phase-leading right-handed magnetic dipole.
 \item \textbf{Linear Huygens Basis Viewpoint:} $(p_x - m_y) + i (p_y + m_x) \rightarrow h_x + i h_y$. 
 
 This shows that the exact same radiation pattern can be understood as the interference of two orthogonal, forward-propagating *linear* Huygens dipoles that are driven 90$^\circ$ out of phase.
 \item \textbf{Chiral Basis Viewpoint:} $(p_x + i m_x) + i (p_y + i m_y) \rightarrow \sigma_x^+ + i \sigma_y^+$. 
 
 This reveals a third perspective: the interference of two orthogonal *chiral* dipoles driven in quadrature.
 
\end{itemize}
This equivalence, which holds for all four unidirectional spin states, is profound. It demonstrates that the concepts of circular, chiral, and Huygens dipoles are not mutually exclusive but are deeply interconnected facets of the same underlying physics. The Circular Huygens Dipole is not just a new type of source, but a concept that unifies these different pictures to achieve a novel functionality: the directional emission of spin-angular momentum.

\section{Multipole Decomposition Formulas}
\label{app:formulas}

The induced electric current density, $\vec{J}(\vec{r})$, is determined from the simulated internal electric field $\vec{E}(\vec{r})$ via the relation $\vec{J} = -i\omega\epsilon_0(\epsilon_r - 1)\vec{E}$ \cite{Evlyukhin2019}, where $\epsilon_r$ is the relative permittivity of silicon. To accurately determine the multipole moments excited in the nanocuboid from this current, we employ the Cartesian expressions derived by Alaee et al.\cite{Alaee2018_SM}. The complete expressions for the Cartesian components ($\alpha, \beta \in \{x,y,z\}$) are provided below. Here, $k$ is the wavenumber and $j_n(kr)$ are spherical Bessel functions.

\textbf{Electric Dipole (ED):}
\begin{align}
 p_{\alpha}=&-\frac{1}{i\omega}\int d^{3}r \bigg\{ J_{\alpha} j_{0}(kr) \nonumber\\
 &+\frac{k^{2}}{2}[3(\vec{r}\cdot \vec{J})r_{\alpha}-r^{2}J_{\alpha}]\frac{j_{2}(kr)}{(kr)^{2}}\bigg\}
\end{align}
\textbf{Magnetic Dipole (MD):}
\begin{equation}
 m_{\alpha}=\frac{3}{2}\int d^{3}r(\vec{r}\times \vec{J})_{\alpha}\frac{j_{1}(kr)}{kr}
\end{equation}
\textbf{Electric Quadrupole (EQ):}
\begin{align}
 Q^e_{\alpha\beta} = &-\frac{3}{i\omega} \bigg\{ \int d^{3}r[3(r_{\beta}J_{\alpha}+r_{\alpha}J_{\beta}) \nonumber\\
 & \qquad -2(\vec{r}\cdot \vec{J})\delta_{\alpha\beta}]\frac{j_{1}(kr)}{kr} \nonumber \\
 & + 2k^{2}\int d^{3}r[5r_{\alpha}r_{\beta}(\vec{r}\cdot \vec{J}) \nonumber\\
 & \qquad -(r_{\alpha}J_{\beta}+r_{\beta}J_{\alpha})r^{2}-r^{2}(\vec{r}\cdot \vec{J})\delta_{\alpha\beta}]\frac{j_{3}(kr)}{(kr)^{3}} \bigg\}
\end{align}
\textbf{Magnetic Quadrupole (MQ):}
\begin{align}
 Q^m_{\alpha\beta}=&15\int d^{3}r\{r_{\alpha}(\vec{r}\times \vec{J})_{\beta} \nonumber\\
 &\qquad +r_{\beta}(\vec{r}\times \vec{J})_{\alpha}\}\frac{j_{2}(kr)}{(kr)^{2}}
\end{align}

\subsection*{Scattering Efficiency}
The spectral plots in Fig.~\ref{fig2}(b) and Fig.~\ref{fig4}(b) show the magnitude of the scattering efficiency contribution from each individual Cartesian multipole moment. The scattering efficiency is defined as the scattering cross-section, $C_{\text{sca}}$, normalized by the geometrical cross-section of the nanoparticle, $A_{\text{geom}} = L_x L_y$. The contribution of each multipole component to the total scattering efficiency is given by the following expressions, where $|E_\text{inc}|$ is the amplitude of the incident electric field:
\begin{align}
 Q_{\text{sca}}(p_\alpha) &= \frac{k^4}{6\pi\epsilon_0^2 |E_\text{inc}|^2 A_{\text{geom}}} |p_\alpha|^2 \\
 Q_{\text{sca}}(m_\alpha) &= \frac{k^4}{6\pi\epsilon_0^2 c^2 |E_\text{inc}|^2 A_{\text{geom}}} |m_\alpha|^2 \\
 Q_{\text{sca}}(Q^e_{\alpha\beta}) &= \frac{k^6}{720\pi\epsilon_0^2 |E_\text{inc}|^2 A_{\text{geom}}} |Q^e_{\alpha\beta}|^2 \\
 Q_{\text{sca}}(Q^m_{\alpha\beta}) &= \frac{k^6}{720\pi\epsilon_0^2 c^2 |E_\text{inc}|^2 A_{\text{geom}}} |Q^m_{\alpha\beta}|^2
\end{align}
The total scattering efficiency is the sum of these individual contributions.


\end{document}